\newcommand{\zlabel}[1]{\label{#1} }
\newcommand{\fc}{\frac} 
\newcommand{\lt}{\left} 
\newcommand{\rt}{\right} 
\newcommand{\mr}{\mathbf{r}}
\newcommand{\pr}{\prime}

\newcommand{\ms}{\mathbf{s}}
\newcommand{\mx}{\mathbf{x}}

\newcommand{\al}{\alpha}

\newcommand{\pa}[2]{\frac{\partial #1}{\partial #2}}

\newcommand{\bu}{\mathbf{u}}

\newcommand{\lsc}{\;\text{\Large ;}}

\documentclass[pra,preprint,preprintnumbers,amsmath,amssymb,floatfix]{revtex4}
\usepackage{graphicx}
\usepackage{dcolumn}
\usepackage{epic}%
\usepackage{eepic}%
\usepackage{amsthm}
\usepackage{color}
\usepackage{bm}
\usepackage{comment}
\usepackage{esvect}
\usepackage{textcomp}

\begin{document}

\title{A Quantum Mechanics Conservation of Energy Equation for Stationary States with Real Valued Wave Functions}

\author{James P. Finley}
\email{james.finley@enmu.edu}
\affiliation{
Department of Physical Sciences,
Eastern New~Mexico University,
Station \#33, Portales, NM 88130}
\date{\today}

\begin{abstract}
Many-body quantum-mechanical stationary states that have real valued wavefunction are shown to
satisfy a classical conservation of energy equation with a kinetic energy function.  The terms
in the equation depend on the probability distribution, and, in addition, pressure and
velocity functions, but these functions also depend on the probability distribution. There are
two possible directions of the velocity that satisfy the energy equation.
A linear momentum function is defined that integrates to zero, and this property is consistent
with the expectation value of the linear momentum for stationary states with real-valued wave
functions.
The energy equation is integrated to obtain a version of the well known energy equation
involving reduced density matrices, where the kinetic energy functional of the one-particle
density matrix is replaced by a function of the electron density and a velocity function.  Also,
the noninteracting kinetic energy functional from the Hohenberg--Kohn theorem is given as an
explicit functional of the orbital densities.
For the purpose of describing the behavior of particles in a stationary state, a model based on
the energy equation is consturcted.
The model is evaluated for the two different velocity directions using the grounds state of the
particle in a one-dimensional box and the hydrogen atom.
For one velocity direction, equations of motions with contradictory properties are obtained,
and, in the other, an unstable system is found.
A discussion is given with suggestions of additional elements that might improve the model.
%
\end{abstract}

\maketitle

\section{Introduction}

There are some correspondence between classical-mechanical states with conservative potentials
and stationary states of quantum mechanics. Both types of states satisfy an equation with the
energy as a constant. For electronic systems, the external potential-energy and
electron-electron repulsion energy functions for the time independent Schr\"odinger equation
are obtained from classical mechanics, and they have a simple interpretation when the electrons
are considered point masses with a negative charge. The correspondence for the kinetic energy
is, for the most part, absent, if the particles are considered point masses with charge.  The
non-classical treatment of the kinetic energy also has consequences for density matrix theory
\cite{Cioslowski,Parr:89}, since the off diagonal elements of the one-particle density matrix, which
seems to have no physical meaning and adds to the complexity of the problem, must be considered.

In this work, many-body quantum-mechanical stationary states that have real valued wavefunction
are shown to satisfy a classical conservation of energy equation with a kinetic energy function
(in Sec.~\ref{p7382}).  The energy equation depends on the probability distribution, i.e., the
square of the wavefunction, and pressure and velocity functions that also depend on the
probability distribution, where there are two possible directions of velocity consistent with
the energy equation.

A linear momentum function (in Appendix~\ref{p3218}) is defined that is consistent with the
kinetic-energy function. As in the expectation value of the total momentum for quantum
mechanical states with real-valued wave functions, the linear momentum function integrates
(over all space and spin variables) to zero. This implies that the linear momentum function can
be neither non-negative nor non-positive, if it is not the zero function.

In a manner that is consistent with the energy equation, a model that contains
classical-mechanical elements is constructed (\ref{p2937}) for the purpose of describing the
behavior of particles in a stationary state. The model is evaluated for the two different
velocity directions using the grounds state of the particle in a one-dimensional box and the
hydrogen atom.  For one velocity direction, equations of motions with contradictory properties
are obtained, and, in the other, an unstable system is found that can only be stabilized by a
nonclassical restriction on the velocity direction. These failures, however, may be useful in
the path to the eventual formulation of a useful model of quantum mechanical states containing
classical mechanical elements that is based on the energy equation.  For example, the
construction of a more accurate model might be possible if the particles can frequently change
velocity directions with no change in speed, or if a nonclassical restriction on the velocity
direction is imposed, i.e., the particle only travels in the opposite direction of a force
field.

The energy equation is integrated (\ref{p5282}) to obtain a version of the well known energy
equation \cite{Lowdin:55,McWeeny:60,Parr:89} involving reduced density matrices, where the
kinetic energy functional of the one-particle density matrix is replaced by a function of the
electron density and a velocity function.  The final result is an equation that does not depend
on the off diagonal elements of a reduced density matrix.

The noninteracting kinetic energy functional $T_s$ from the Hohenberg--Kohn
theorem \cite{Parr:89,Dreizler} is given as an explicit functional of the orbital densities (\ref{p2952}).


\section{The quantum mechanics conservation of energy equation for stationary states \zlabel{p7382}} 

  

The $n$-body time-independent Schr\"odinger equation with a normalized, real-valued eigenfunction $\psi$
can be written
\begin{equation} \zlabel{p2602}
  -\fc{\hbar^2}{2m}\sum_{i=1}^n \lt[\psi\nabla_i^2\psi\rt] + \sum_{i=1}^n v_i\gamma_n + \fc12 \sum_{i\ne j}^n r_{ij}^{-1}\gamma_n = \bar{E}\gamma_n
\end{equation}
where 
\[
\lt[\psi\nabla_i^2\psi\rt](\mx) = \psi(\mx)\nabla_{\mr_i}^2\psi(\mx), \quad \mx = \mathbf{x}_1,\cdots \mathbf{x}_n
\]
Also, the $n$-body probability distribution $\gamma_n$---the \emph{diagonal} part of the $n$-body
density matrix---is $\gamma_n = \psi^2$; the electron coordinate $\mathbf{x}_i$ is
defined by $\mathbf{x}_i = \mathbf{r}_i,\omega_i$, where $\mathbf{r}_i\in\mathbb{R}^3$ and $\omega_i\in\{-1,1\}$ are the
spatial and spin coordinates, respectively. Furthermore, the $v_i$ and $r_{ij}^{-1}$
multiplicative operators are defined by the following:
\[
[v_i\gamma_n](\mx) =  v(\mr_i)\gamma_n(\mx), \quad [r_{ij}^{-1}\gamma_n](\mx) = |\mr_i - \mr_j|^{-1}\gamma_n(\mx) \lsc  
\]
where the one-body external potential $v$ is a specified real-valued function with domain
$\mathbb{R}^3$ such that $\{\mr\in\mathbb{R}^3| \psi(\mx) = 0\}$ has measure zero. This last
requirements for $v$ implies that the division of a equation by $\psi$ or $\gamma$ gives an
equation that is defined almost everywhere.  Henceforth, to reduce cluttter, the $n$
subscript appearing in the symbol $\gamma_n$ is suppressed.

Substituting the following identity
\begin{equation} \zlabel{0152}
-\fc12\lt[\psi\nabla_i^2\psi\rt] = \fc18\lt[\gamma^{-1}\nabla_i\gamma\cdot\nabla_i\gamma\rt] - \fc14\nabla_i^2\gamma,
\end{equation}
into the Schr\"odinger' (\ref{p2602}), where the identity is easily derived \cite{Finley1}, gives
\[
\fc{\hbar^2}{8m}\sum_{i=1}^n\gamma^{-1} \lt|\nabla_i\gamma\rt|^2- \fc{\hbar^2}{4m}\sum_{i=1}^n\nabla_i^2\gamma  +
  \sum_{i=1}^n v_i\gamma + \fc12 \sum_{i\ne j}^n r_{ij}^{-1}\gamma = \bar{E}\gamma
\]
Using the definitions
\begin{equation} \zlabel{p4720}
\bu_{i\pm} = \pm\fc{\hbar}{2m}\fc{\nabla_i\gamma}{\gamma}, \qquad  p_i = -\fc{\hbar^2}{4m}\nabla_i^2\gamma
\end{equation}
and
\[
\fc12 m u_i^2 = \fc12 m\bu_{i\pm}\cdot\bu_{i\pm} = \fc12 m \lt|\fc{\hbar}{2m}\fc{\nabla_i\gamma}{\gamma}\rt|^2
= \fc{\hbar^2}{8m} \gamma^{-2} \lt|\nabla_i\gamma\rt|^2
\]
we have
\begin{equation} \zlabel{p4791}
\sum_i \fc12 m \gamma u_i^2  + \sum_i p_i +   \lt(\sum_{i=1}^n v_i\rt)\gamma + \lt(\fc12 \sum_{i\ne j}^nr_{ij}^{-1}\rt)\gamma = \bar{E}\gamma,
\end{equation}

For any point $\mx$ such that $\gamma(\mx) \ne 0$, this equation can also be written
\begin{equation} \zlabel{p4791b}
\sum_i \fc12 m u_i^2  + \lt(\sum_i p_i \rt) \gamma^{-1} +   \sum_{i=1}^n v_i + \fc12 \sum_{i\ne j}^nr_{ij}^{-1} = \bar{E}
\end{equation}
Another way to write it is 
\begin{equation} \zlabel{p4792}
    T + P\gamma^{-1} + V_{\text{nuc}} + V_{\text{ee}} = \bar{E}
\end{equation}
where
\[
T = \sum_i^n \fc12m u_i^2, \quad P = \sum_i^n p_i, \quad  V_{\text{nuc}} = \sum_{i=1}^n v_i, \quad  V_{\text{ee}} = \fc12 \sum_{i\ne j}^nr_{ij}^{-1}
\]
and where $P$ is called the pressure and $P\gamma^{-1}$ is called the compression energy, explicitly defined by
\begin{gather} \notag
  [P\gamma^{-1}](\mx) = \gamma^{-1}(\mx)\sum_i^n \lt(-\fc{\hbar^2}{4m}\nabla_{\mr_i}^2\gamma(\mr_1,\omega_1;\cdots \mr_n,\omega_n) \rt)
\end{gather}
Eq.~(\ref{p4792}) is a quantum-mechanical classical equation of the conservation of energy $\bar{E}$.

Next consider a one-body state, such that $\psi(\mr,1) = \phi(\mr)\alpha(1)$, where $\alpha$ is
the spin function that satisfies $\alpha(1)= 1$ and $\alpha(-1)= 0$. Hence, $\gamma(\mr,1) =
\phi^2(\mr)\;\dot{=}\;\rho(\mr)$, and the last ``equality'' is a definition of the electron
density $\rho$.  In this special case, (\ref{p4720}) and (\ref{p4791b}) can be written
\begin{gather}
\zlabel{p2572}  \bu_{\pm} = \pm\fc{\hbar}{2m}\fc{\nabla\rho}{\rho}, \quad p = \fc{\hbar^2}{4m}\nabla^2\rho \\
\zlabel{p2574}  \fc12 m u^2 + p\rho^{-1} + v = \bar{E} 
\end{gather}
This last equation has been used to as a starting point to treat one-body steady-state quantum
systems with real valued wave functions as fluids satisfying variants of well know equations
from fluid dynamics, with body force $v$, pressure $p$, velocity field $\bu_\pm$ and mass
density~$m\rho$ \cite{Finley1}.

\section{A model consistent with the energy equation (\ref{p4792}) \zlabel{p2937}}

Since we have an energy equation (\ref{p4792}), a reasonable next step is the description of
states of systems that satisfy specified constraints, such that (\ref{p4792}) is satisfied.
Another words, one or more models are needed. In this section, a model is given where the
electrons are point masses with charge, and they behave like classical particles.

\subsection{Components of a  model}

Let $(\mr, \mr + d\mr)$ denote a region of $\mathbb{R}^3$, given by
\begin{gather*}
  (\mr, \mr + d\mr) \, \dot{=} \, (x, x + dx)\times(y, y + dy)\times(z, z + dz) \in \mathbb{R}^3 
\end{gather*}
where we use `$\dot{=}$' for definitions.  Let $\gamma$ represent a stationary state such that
the values of the spin variables and the locations of $n$ electrons can be determined by
experiment.  The well known meaning given to $\gamma$ is the following
\cite{Lowdin:55,McWeeny:60,Parr:89}: For a state represented by $\gamma$, the real non-negative
number
\begin{equation} \zlabel{p7320}
 \bar{\gamma} = 
 n!\gamma(\mx)\, d\mr_1\, d\mr_2\cdots d\mr_n; \qquad
 \mx = \mr_1,\omega_1; \mr_2,\omega_2, \cdots \mr_n,\omega_n,
\end{equation}
is the probability of \emph{finding} an electron (any one) in the volume element $(\mr_1, \mr_1
+ d\mr_1)$ with spin variable $\omega_1$, an electron (a different one) in the volume element
$(\mr_2, \mr_2 + d\mr_2)$ with spin variable $\omega_2$, $\cdots$ and an electron (the last
one) in the volume element $(\mr_n, \mr_n + d\mr_n)$ with spin variable $\omega_n$.  In other
words, it is the probability of finding the $n$ electrons, in any order, in the following
sequence of combined volume elements and spin variables: $(\mr_1, \mr_1 + d\mr_1)$, $\omega_1$;
$(\mr_2, \mr_2 + d\mr_2)$, $\omega_2$; $\cdots$ $(\mr_n, \mr_n + d\mr_n)$, $\omega_n$. Note
that $\bar{\gamma}$ is the value of a function at the point $\mx$ and $n$ regions of $\mathbb{R}^3$
that is now specified.

Next we present a model that give interpretations for the functions from the previous section.
First, however, a different meaning is given to $\bar{\gamma}$, obtained by replacing the words
'the probability of finding' in the above interpretation with 'the fraction of time:'
\begin{quote} {\bf I)}
  The real non-negative number $\bar{\gamma}$ is the fraction of time the $n$ electrons, in any
  order, are contained in the following volume elements and have the following spin variables:
\newline
\mbox{}\hspace{5ex}$(\mr_1, \mr_1 + d\mr_1)$, $\omega_1$; $(\mr_2, \mr_2 + d\mr_2)$, $\omega_2$; $\cdots$
  $(\mr_n, \mr_n + d\mr_n)$, $\omega_n$.
\end{quote}
Using interpretation {\bf I} the following definition has a meaning without mentioning an
experimental measurement:

{\bf Definition.} An $n$-electron state is said to be at the spatial-spin location $\mathbf{x}=
\{\mr_1,\omega_1, \cdots \mr_1,\omega_1\}$ if there is an electron (any one) at the coordinate
$\mr_1$ with spin variable $\omega_1$, an electron (a different one) at the coordinate $\mr_2$
with spin variable $\omega_2$, $\cdots$ and an electron (the last one) at the coordinate
$\mr_n$ with spin variable $\omega_n$.\vspace{1ex}

Next we give an interpretation to the energy terms in (\ref{p4792}) and the pressure $P$.
\begin{quote}
Let the $n$-electron state be at the spatial-spin location $\mathbf{x}$ such that $\gamma(\mathbf{x}) \ne 0$.

{\bf IIa)} The functions $T(\mx)$, $[P\gamma^{-1}](\mx)$, $V_{\text{nuc}}(\mr)$,
$V_{\text{ee}}(\mr)$ and $P(\mx)$ are the total values of the kinetic- $T$, compression-
$P\gamma^{-1}$, external potential- $V_{\text{nuc}}$, and the electron-electron repulsion-
$V_{\text{ee}}$ energies; and, $P(\mx)$ is the total value of the pressure.
\end{quote}
Note that, all things being equal, the compression energy $P\gamma^{-1}$ is relatively large
for spin-spatial locations $\mathbf{x}$ that have a small fraction of time $\bar{\gamma}$.

Next we give an interpretation to individual functions that contribute to the energy terms 
in (\ref{p4792}).
\begin{quote}
{\bf IIb)} The function $v(\mr_i)$ is external potential for the electron located at
$\{\mr_i,\omega_i\}$ from \emph{any} spatial-spin location $\mathbf{x}$ containing
$\{\mr_i,\omega_i\}$. The function $r_{ij}^{-1}$ is the electron-electron repulsion energies
between the electrons at $\{\mr_i,\omega_i\}$ and $\{\mr_j,\omega_j\}$ from \emph{any} spatial-spin
location $\mx$ containing $\{\mr_i,\omega_i\}$ and $\{\mr_j,\omega_j\}$.
The functions $\bu_{\pm i}(\mx)$ and $p_i(\mx)$ are the velocity and pressure for the electron
located at $\{\mr_i,\omega_i\}$ from \emph{the} spatial-spin location $\mathbf{x}$ that
contains $\{\mr_i,\omega_i\}$, where there are two possible velocity directions, called uphill
$\bu_{+i}$ and downhill $\bu_{-i}$ velocity.
\end{quote}

\subsection{Reservations and discussion}
  
Interpretation {\bf IIb)} might contradict the uncertainty principle for position and momentum, because
when the $n$-electron state is at the spatial-spin location $\mathbf{x}$, the momentum is either
$m\bu_{+i}(\mx)$ or $m\bu_{-i}(\mx)$ for the electron located at $\{\mr_i,\omega_i\}$.

For this paragraph, suppose {\bf IIb} violates the uncertainty principle. This does not imply
that a useful model to explain a class of phenomena cannot be constructed if it contains {\bf
  IIb}, since useful models with flaws are common. One example is the assignment of orbital
energies to a special class of orbitals that define a Slater single-determinantal  wavefunction
of Hartree--Fock theory. However, there exist other orbital sets that can define the same
determinantal wavefunction, and these orbitals cannot be assigned orbital energies, at least not
in the same way.  Also, it is useful to have a simple starting point, where, later on, it might
be possible to add corrections to treat this deficiency. For example, a Hartree product
wavefunction does not provide an anti-symmetric wavefunction and distinguishes electrons, but
it is an excellent starting point to introduce Slater-determinantal wavefunctions, since
the individual terms of a Slater-determinant are Hartree products.

Let $\phi$ be a normalized wavefunction.  Let $\chi$ be a normalized, nondegenerate eigenvector
with eigenvalue $\lambda$ of a Hermitian operator $\hat{O}$ with a discrete spectrum
representing an observable, such that the inner product $(\phi,\chi)$ of the $L^2$ Hilbert
space satisfies $|(\phi,\chi)|^2\in (0,1)$. Hence, $\phi$ and $\chi$ are linerly independent,
but not orthogonal. According to an axiom of quantum mechanics \cite{Byron,Jordon}, if a
measurement is made of the observable with operator $\hat{O}$ of the state represented by
$\phi$, then the probability of the measured value being equal to $\lambda$ is
$|(\chi,\phi)|^2$. Also, if the measured value is $\lambda$, then the quantum mechanical system
is in the state $\chi$ immediately after the measurement is made. In other words, {\em the
  measurement has transformed the state $\phi$ into $\chi$}. The same type of dramatic
transformations are easily shown to hold for observers represented by bounded self-adjoint
operators such that $\lambda$ corresponds to a degenerate subspace, and to cases where the operator
$\hat{O}$ is unbounded, but still self-adjoint with a continuous spectrum.  If it is possible
to supplement the axions of quantum mechanics by adding measurements that are not, in general,
so destructive, but still provide information, then the interpretations above seem more
reasonable. In a thought experiment it is easy to imagine the electron as having a specific
position and momentum before a ``measurement of destruction'' is made, as in a position
measurement that traps the electron in an infinitesimal three dimensional box.


\subsection{Testing the model with velocity restrictions}

Next, by applying (\ref{p2572}) and (\ref{p2574}) to two one-body states, we expose some
problems with interpretations {\bf I} and {\bf II}, if the particle has a classical equation of
motion that contains one and only one velocity function, that is, the single particle is
described by either uphill $\bu_{+}$ and downhill $\bu_{-}$ velocity.  We will also consider
what additional restrictions or changes can be made to improve the situation. The calculations
we do below have already been done for another purpose, where (\ref{p2572}) and (\ref{p2574})
are used to describe the fluid dynamics of one-body quantum states \cite{Finley1}, where other
one-body states are also treated.

\subsubsection{Equations of motion with downhill velocity} 

For a particle in a one-dimensional box with a length that is 1 Bohr radius $a_0$, where the well known
probability density in atomic units is $\rho(r) = 2\sin^2(n\pi r)$ \cite{Raimes,Bransden}, it is easily demonstrated
that \cite{Finley1}
\begin{gather*}
u_\pm(r) = \pm \fc12\rho^{-1}(r)\fc{\partial\rho(r)}{\partial r} = \pm n\pi\cot(n\pi r), \quad n = 1,2,\cdots \\
p(r) = -n^2\pi^2[1 - 2\sin^2(n\pi r)]
\end{gather*}
for the two possible velocity directions, where $r\in [0,1]$. Fig.~\ref{box1-energy} gives a
classical energy diagram that also contains the downhill velocity $u_-$.  Such diagrams of the
potential and total energies are a familiar part of classical mechanics \cite{Marion}.  Since
there is no external potential, the compression energy $p\rho^{-1}$ is the potential energy.
Hence the distance between $\bar{E}_1$ and $p\rho^{-1}$ in the diagram is equal to the kinetic
energy, as indicated by the energy equation~(\ref{p2574}).  For the ground state ($n =1$) the
speed is zero at $r = 1/2$ and infinity at the nodes (at $r = 0,1$). The point $r = 1/2$ is an
unstable equilibrium point.  A particle with downhill velocity $u_-$ at the point $r = 1/2 +
\epsilon$, where $ 0 < \epsilon < 1/2$, will accelerate to the right and reach $r = 1$ in a
finite amount of time, where the velocity is infinity, and where, according to interpretation
{\bf I}, the particle spends no time, since $r = 1$ is a node. Since, $u_{-}(0.5 - \epsilon) =
-u_{-}(0.5 + \epsilon)$, the same behavior is exhibited for $r\in [0.1/2]$. Hence, a necessary
condition for interpretations {\bf II} to provide a useful model is that the particle must
frequently change directions with constant speed, or transport---be destroyed at one point in space and
simultaneously be created in another---and, if possible, do so in a way such that
interpretation {\bf I} is satisfied.
 \begin{figure}[!tbp] \input{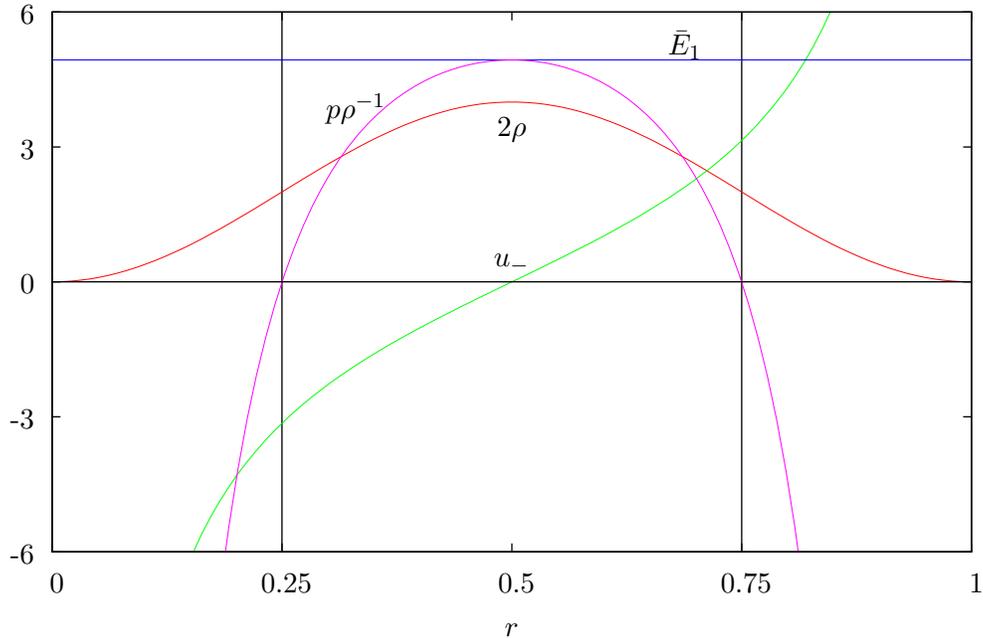} \caption{The compression $p\rho^{-1}$ and total $\bar{E}_1$ energies with the probability density $\rho$
  and downhill velocity $u_-$ of the ground-state of a particle in a one-dimensional box of length $a_0$
  in atomic units from \cite{Finley1}. \zlabel{box1-energy}} \end{figure}

For one electron atoms in the 1s state with atomic number $Z$ in spherical coordinates we have
\[
\rho(r) = \fc{Z^3}{a_0^3}\pi^{-1}e^{-2Zr/a_0}, \qquad \bu_\pm = \mp \fc{Z\hbar}{ma_0}\hat{\mr}
\]
where the density is well known \cite{Raimes,Bransden}, and the velocity $\bu_\pm$ is compued
using (\ref{p2572}).  (The pressure is examined in detail elsewhere for the hydrogen atom
\cite{Finley1}.)  For $Z = 1$, $|\bu_\pm| = \hbar/ma_0$, and this is the speed of the electron
in the first Bohr orbit of hydrogen \cite{Bransden}. Hence, the speed is constant and the
electron is restricted to a single ray, or two opposite rays, if the electron can pass through
the nucleus; the electron has zero angular momentum.  For downhill velocity, the same necessary
condition mentioned for the particle in a box is applicable for these states, except that the
electron also must be able to change rays that are not on opposite sides of the nucleus.

{\bf Conclusion \#1.} The cost of violating the uncertainty principle is a particle that
in a finite amount of time has infinite speed and is traveling no where, or one
that does nothing more than travel towards an infinity.

{\bf Conclusion \#2.}  According to Einstein ``Everything should be made as simple as possible,
but not simpler.''  Perhaps the proposed model is too simple.  In order for the above
interpretations to be improved to give a satisfactory model, the interpretations must be
modified or additional rules or variables must be added, and these variable(s) might need to be
random. Also, since the states have pressure, it might be necessary for the electron to have a
nonzero volume and be compressible.

\subsubsection{Equations of motion with uphill velocity}

For the particle in a box, if the particle is given an initial uphill velocity $u_+$ at a point
$r = 1/2 + \epsilon$, the particle moves in the direction of increasing electron density
$\rho$ until it becomes ``perched'' at $r = 1/2$ in an infinite amount of time, because the speed of
the particle is zero at $r = 1/2$.  However, stability is achieved at the point $r = 1/2$ only
if the particle is restricted to uphill velocity $u_+$, so it cannot be `pushed off the top of
the density hill,' so to speak, at $r = 1/2$. This constraint does not have a classical
correspondence.

For one electron atoms in the 1s state, there is a velocity discontinuity at the nucleus $r =
0$. It is difficult to see how this discontinuity can be consistent with an equation of motion
for an electron described as a point-mass particle. However, if the external electrostatic
potential is modified to take into account mass and volume effects of the nucleus, as in
isotope shifts \cite{Bransden}, then the external potential should have a minimum at the
nucleus. Hence, the gradient of the electron density as $r \to 0$ is zero, giving a zero speed
at the nucleus.  This case is then modeled very well by the particle in a box, and with the
same reservations.

{\bf Conclusion.} Uphill velocity is only stable if the particle cannot have downhill velocity,
and this restriction has no correspondence with classical states.  The restriction implies
that the particle only translates in the opposite direction of a force field.

\section{The density matrix equation without off diagonal elements \zlabel{p5282}}

Integrating (\ref{p2602}) over all space and summing over all spin variables in atomic units,
we obtain the well-known density-matrix energy equation \cite{Lowdin:55,McWeeny:60,Parr:89}:
\begin{equation} \zlabel{p2127}
 \bar{E} =  \fc{\hbar}{2m}\int\lt[\nabla_\mr^2\rho_1(\mr,\mr^\pr)\rt]_{\mr^\pr=\mr} \,d\mr
  + \int v(\mr)\rho(\mr) \,d\mr + \int\int r_{12}^{-1}\rho_2(\mr_1,\mr_2)\,d\mr_1 \,d\mr_2\lsc
\end{equation}
the one-body density matrix $\rho_1$ and pair function $\rho_2$ are defined by 
\begin{align*}
\rho_1(\mathbf{r}_1,\mathbf{r}_1^\pr) =&\,
n \sum_{\omega_1\cdots\omega_n} \int \psi(\mr_1,\omega_1,\mathbf{x}_2,\mathbf{x}_3,\cdots \mathbf{x}_n) 
\psi(\mr_1^\pr,\omega_1,\mathbf{x}_2,\mathbf{x}_3,\cdots \mathbf{x}_n) 
\, d\mathbf{r}_2 d\mathbf{r}_3 \cdots d\mathbf{r}_n,
\\
\rho_2(\mathbf{r}_1,\mathbf{r}_2) =&\,
\fc{n(n-1)}{2} \sum_{\omega_1\cdots\omega_n} \int \psi^2
\, d\mathbf{r}_3 d\mathbf{r}_4\cdots d\mathbf{r}_n,
\end{align*}
and the electron coordinate $\mathbf{x}_i$ is defined by $\mathbf{x}_i = \mathbf{r}_i,\omega_i$, where $\mathbf{r}_i$
and $\omega_i$ are the spatial and spin coordinates, respectively; the one-body density $\rho$ is defined by
$\rho(\mr) = \rho_1(\mr,\mr)$.


Our objective is to integrate (\ref{p4791}) over all space and sum over all spin coordinates
and obtain an expression that resembles (\ref{p2127}).
Let $\mr_i$ be a Cartesian coordinate, such that $\mr_i = (x_i,y_i,z_i)$ and
$\al_i\in(x_i,y_i,z_i)$. We require that $\psi$ satisfies the usual boundary conditions:
\[
\lim_{\al_i \to \infty} \psi(\mx) = \lim_{\al_i \to \infty} \pa{\psi(\mx)}{\al_i} = 0
\]
Using this requirement and $\partial \gamma /\partial \al_i = 2\psi (\partial \psi/\partial\al_i)$ we have
\begin{gather*}
  \int_{-\infty}^{+\infty} \fc{\partial}{\partial \al_i}\lt(\fc{\partial\gamma }{\partial \al_i}\rt)  \,d\al_i =
  2\lt(\lim_{\al_i \to \infty} - \lim_{\al_i \to -\infty}\rt)\psi\fc{\partial\psi}{\partial \al_i} = 0 
\end{gather*}
Using this identity and definition (\ref{p4720}) it is easily proved that 
\[
\int p_i(\mx) \,d\mr_i = 0 
\] 
Hence, the second term from (\ref{p4791}) will not contribute after integrating over all space.
Also, the pressure function $p_i$ can be neither non-negative nor non-positive, if it is not the zero
function.

Let the nonnegative function $\bar{u}_1^2$ with domain $\mathbb{R}^3- \{\mr\in\mathbb{R}^2| \psi(\mr) = 0\}$ be defined by
\begin{gather*}
\rho(\mr_1)\bar{u}_1^2(\mr_1) = n\!\!\sum_{\omega_1\cdots\omega_n} \int \gamma(\ms) u_1^2(\ms) 
\, d\mathbf{r}_2 \cdots d\mathbf{r}_n
\end{gather*}
Using this definition, we treat the first term from (\ref{p4791}):
\begin{gather*}
    \fc12 \sum_i \sum_{\omega_1\cdots\omega_n} \int m\gamma(\ms)  u_i^2(\ms)\, d\mathbf{r}_1 \cdots d\mathbf{r}_n 
\hspace{40ex} \\
  = \fc12 n\!\!\sum_{\omega_1\cdots\omega_n} \int m \gamma(\ms) u_1^2(\ms) 
\, d\mathbf{r}_1 \cdots d\mathbf{r}_n
 = \fc12 \int \rho_m(\mr_1)\bar{u}_1^2(\mr_1)\, d\mr,
\end{gather*}
where $\rho_m = m\rho$.  Using this result and the result above for the integration of the
pressure term, we can integrate (\ref{p4791}) over all space and sum over all spin coordinates
to obtain our objective:
\begin{gather}
  \bar{E} = \fc12 \int \rho_m(\mr)\bar{u}^2(\mr)\, d\mr
  + \int v(\mr)\rho(\mr) \,d\mr + \int\int r_{12}^{-1}\rho_2(\mr_1,\mr_2)\,d\mr_1 \,d\mr_2
\end{gather}
The classical electrostatic interpretation of the second term is that it is the electrostatic
energy assigned to the charge density $q\rho$, also called a charge cloud, where $q$ is the charge of the
indistinguishable particles. Similarly, the first term is the kinetic energy of the mass
density or ``mass cloud,'' a continuum in motion, i.e., a fluid, where $\rho_m(\mr)$ and $|\bar{u}(\mr)|$
are the average mass density and speed at $\mr$, respectively.

\section{The kinetic energy functional of the Hohenberg--Kohn theorem of noninteracting electrons \zlabel{p2952}}

Equation~(\ref{0152}), in the fictitious case of a single electron without spin, is
\[
-\fc12\phi\nabla^2\phi = \fc18\rho^{-1}\nabla\rho\cdot\nabla\rho - \fc14\nabla^2\rho,
\]
where $\rho = \phi^2$, and this equation also holds for a case where a value of the spin
variable is taken, as in (\ref{p2572}) and (\ref{p2574}).  Using this equation, the kinetic
energy functional $K_s$ from the Hohenberg--Kohn theorem \cite{Parr:89,Dreizler} for the
special case of \emph{one-electron states} with a non-degenerate wavefunction $\psi$ is,
explicitly given by
\begin{equation} \zlabel{p7128}
  K_s[\rho] = -\fc{\hbar^2}{2m}\int_{\mathbb{R}^3} \phi\nabla^2\phi\, d\mr
  = \fc{\hbar^2}{m}\int_{\mathbb{R}^3} \lt(\fc18\rho^{-1}\nabla\rho\cdot\nabla\rho - \fc14\nabla^2\rho\rt)\,d\mr,
\end{equation}
and note that any non-degenerate ground state can be represented by a real valued eigenfunction.
%
%
Eq.~(\ref{p7128}) is an equality, not an identity, that holds for the set of all one-electron
stationary states such that $\rho$ is either the electron density of a nondegenerate ground-state or a
degenerate state that has a real-valued eigenfunction.

Consider a ground state of $N$ noninteracting electrons, with spin not included,
and with the Hamiltonian
\[
\hat{H} = \sum_{i=1}^N \hat{h}_i, \qquad \hat{h}_i = -\fc12\nabla^2\phi(\mr_i) + v_s(\mr),
\]
where the ground-state eigenfunction is a \emph{real-valued} Slater determinant constructed
from $N$ real-valued, orthonormal, spatial-orbitals $\{\phi_1,\cdots \phi_N\}$.  Using
(\ref{p7128}), the well know noninteracting kinetic energy functional $T_s$ from the
Hohenberg--Kohn theorem \cite{Parr:89,Dreizler} is given as an explicit functional of the
orbital densities $\rho_i = \phi_i^2$:
\begin{gather*}
  T_{s}[\rho]  = -\fc{\hbar^2}{2m}\sum_{i=1}^N \int_{\mathbb{R}^3} \phi_i\nabla^2\phi_i\, d\mr =  \sum_{i=1}^N  K_s[\rho_i]\lsc
  \qquad
\rho = \sum_{i=1}^N \rho_i,
\end{gather*}
and this functional is a key component of the Kohn--Sham methodology of density functional theory \cite{Parr:89,Dreizler}.

\appendix

\section{The linear momentum vector function \zlabel{p3218}}



In this section a total linear momentum function is defined that integrates over all space to
zero, and this property is consistent with the expectation value of the linear momentum for
stationary states with real-valued wave functions.

For any functions $Y = Y(\mx)$ such that $\gamma \mapsto Y$, i.e., $\gamma$ determines $Y$, let
$\langle Y \rangle_\gamma$ be defined by
\[
\langle Y \rangle_\gamma =  \sum_{\omega_1\cdots\omega_n}\int Y(\mx)\; d\mr_1 \cdots d\mr_n 
\]
where $\gamma = \phi^2$ and $\phi = \phi(\mx)$ is a real-valued function that is a member of
the $\mathbf{L}^2$ Hilbert space.

Let $\text{I}_{\mathbb{R}}$ be an index set of three symbols, given by $\text{I}_{\mathbb{R}} =
\{x,y,z\}$.  Let $\hat{L}_\al$ denote the $\al$ component of the total linear momentum operator
$\hat{L}$, i.e.,
\[
\hat{L}_\al = \sum_{j=1}^ni\hbar \pa{}{\al_j}, \quad \al \in \text{I}_{\mathbb{R}}
\]
Since, $\al_j\partial\phi/\partial \al_j$ is real, the expectation value
$\langle\phi|\hat{L}_\al|\phi\rangle$ is pure imaginary or zero.  Since $\hat{L}_\al$ is Hermitian,
any expectation value must be real. Hence, using Dirac notation
\begin{equation} \zlabel{p1530}
\langle\phi|\hat{L}_\al|\phi\rangle = 0.
\end{equation}

Let the total linear-momentum vector-function $\bar{P}_\pm$ be defined by
\begin{equation} \zlabel{p5250}
\bar{P}_\pm = \sum_{i=1}^n m\bu_{i\pm}
\end{equation}
Let $mu_{\al_i \pm}$ denote the $\al\in(x,y,z)$ Cartesian component of $m\bu_{i\pm}$, i.e.,
\[
 mu_{\al_i \pm} \;\dot{=}\; (m\bu_{i\pm})\cdot\hat{\mathbf{e}}_\al = \pm \fc{\hbar}{2} \gamma^{-1}\pa{\gamma}{\al_i}
  = \pm \fc{\hbar}{2} \pa{(\ln\gamma)}{\al_i},
\]
where (\ref{p4720}) is used, and so that $\ln\gamma$ is defined, we use atomic units.  Using
the resolution of the identity, $\bar{P}_\pm$ can be resolved into components
\[
\bar{P}_\pm = \sum_{i=1}^n \sum_{\al\in\text{I}_{\mathbb{R}}}\lt[(m\bu_{i\pm})\cdot\hat{\mathbf{e}}_\al\rt] \hat{\mathbf{e}}_\al= 
\sum_{i=1}^n \sum_{\al\in\text{I}_{\mathbb{R}}}mu_{\al_i \pm} \hat{\mathbf{e}}_\al \\
\]
Hence $\gamma \mapsto \bar{P}_\pm$. From
\[
\int_{-\infty}^{+\infty} mu_{\al_i \pm}(\mx)\; d\al_i
= \pm \fc{\hbar}{2}\lt(\lim_{\al_i\to+\infty}\ln\gamma(\mx) - \lim_{\al_i\to-\infty}\ln\gamma(\mx)\rt) = 0\\
\]
it follows that
\[
  \langle mu_{\al_i \pm} \rangle_\gamma = \sum_{\omega_1\cdots\omega_n}\int mu_{\al_i \pm}(\mx)\; d\mr_1 \cdots d\mr_1  = 0 \\
\]
Hence  
\begin{gather*}
  \langle\phi|\hat{L}|\phi\rangle = \langle \bar{P}_\pm \rangle_\gamma = 0, \qquad \gamma = \phi^2
\end{gather*}
where (\ref{p1530}) is used and
\[
\hat{L} = \sum_{\al\in\text{I}_{\mathbb{R}}}\hat{L}_\al\hat{\mathbf{e}}_\al
\]
is the total linear momentum operator.

The equality $\langle \bar{P}_\pm \rangle_\gamma = 0$ implies that the momentum function can be
neither non-negative nor non-positive, if it is not the zero function, and it is consistent
with $\langle\phi|\hat{L}|\phi\rangle = 0$, the average value of the momentum.


\bibliography{ref}

\begin{thebibliography}{11}
\expandafter\ifx\csname natexlab\endcsname\relax\def\natexlab#1{#1}\fi
\expandafter\ifx\csname bibnamefont\endcsname\relax
  \def\bibnamefont#1{#1}\fi
\expandafter\ifx\csname bibfnamefont\endcsname\relax
  \def\bibfnamefont#1{#1}\fi
\expandafter\ifx\csname citenamefont\endcsname\relax
  \def\citenamefont#1{#1}\fi
\expandafter\ifx\csname url\endcsname\relax
  \def\url#1{\texttt{#1}}\fi
\expandafter\ifx\csname urlprefix\endcsname\relax\def\urlprefix{URL }\fi
\providecommand{\bibinfo}[2]{#2}
\providecommand{\eprint}[2][]{\url{#2}}

\bibitem[{\citenamefont{Cioslowski}(2000)}]{Cioslowski}
\bibinfo{author}{\bibfnamefont{J.}~\bibnamefont{Cioslowski}},
  \emph{\bibinfo{title}{Many-Electron Densities and Reduced Density Matrices}}
  (\bibinfo{publisher}{Kluwer}, \bibinfo{address}{New York, Boston, London},
  \bibinfo{year}{2000}).

\bibitem[{\citenamefont{{R. G. Parr} and Yang}(1989)}]{Parr:89}
\bibinfo{author}{\bibnamefont{{R. G. Parr}}} \bibnamefont{and}
  \bibinfo{author}{\bibfnamefont{W.}~\bibnamefont{Yang}},
  \emph{\bibinfo{title}{Density Functional Theory of Atoms and Molecules}}
  (\bibinfo{publisher}{Oxford University Press}, \bibinfo{address}{Oxford},
  \bibinfo{year}{1989}).

\bibitem[{\citenamefont{L\"owdin}(1955)}]{Lowdin:55}
\bibinfo{author}{\bibfnamefont{P.-O.} \bibnamefont{L\"owdin}},
  \bibinfo{journal}{Phys. Rev.} \textbf{\bibinfo{volume}{97}},
  \bibinfo{pages}{1474} (\bibinfo{year}{1955}).

\bibitem[{\citenamefont{McWeeny}(1960)}]{McWeeny:60}
\bibinfo{author}{\bibfnamefont{R.}~\bibnamefont{McWeeny}},
  \bibinfo{journal}{Rev. Mod. Phys.} \textbf{\bibinfo{volume}{32}},
  \bibinfo{pages}{335} (\bibinfo{year}{1960}).

\bibitem[{\citenamefont{Dreizler and {E.K.U. Gross}}(1990)}]{Dreizler}
\bibinfo{author}{\bibfnamefont{R.~M.} \bibnamefont{Dreizler}} \bibnamefont{and}
  \bibinfo{author}{\bibnamefont{{E.K.U. Gross}}}, \emph{\bibinfo{title}{Density
  Functional Theory}} (\bibinfo{publisher}{Springer-Verlag},
  \bibinfo{address}{Berlin, New York}, \bibinfo{year}{1990}).

\bibitem[{\citenamefont{Finley}(2021)}]{Finley1}
\bibinfo{author}{\bibfnamefont{J.~P.} \bibnamefont{Finley}}
  (\bibinfo{year}{2021}), \bibinfo{note}{posted on http://arXiv.org and to be
  published}.

\bibitem[{\citenamefont{{F. W. Byron, Jr.} and Fuller}(1969)}]{Byron}
\bibinfo{author}{\bibnamefont{{F. W. Byron, Jr.}}} \bibnamefont{and}
  \bibinfo{author}{\bibfnamefont{R.~W.} \bibnamefont{Fuller}},
  \emph{\bibinfo{title}{Mathematics of Classical and Quantum Physics}}
  (\bibinfo{publisher}{Dover}, \bibinfo{address}{New York},
  \bibinfo{year}{1969}).

\bibitem[{\citenamefont{Jordan}(1969)}]{Jordon}
\bibinfo{author}{\bibfnamefont{T.~F.} \bibnamefont{Jordan}},
  \emph{\bibinfo{title}{Linear Opertors for Quantum Mechanics}}
  (\bibinfo{publisher}{Wiley}, \bibinfo{address}{New York, London, Sydney,
  Toronto}, \bibinfo{year}{1969}).

\bibitem[{\citenamefont{Raimes}(1961)}]{Raimes}
\bibinfo{author}{\bibfnamefont{S.}~\bibnamefont{Raimes}},
  \emph{\bibinfo{title}{The Wave Mechanics of Electrons in Metals}}
  (\bibinfo{publisher}{North--Holland}, \bibinfo{address}{Amsterdam},
  \bibinfo{year}{1961}).

\bibitem[{\citenamefont{Bransden and Joachain}(1989)}]{Bransden}
\bibinfo{author}{\bibfnamefont{B.~H.} \bibnamefont{Bransden}} \bibnamefont{and}
  \bibinfo{author}{\bibfnamefont{C.~J.} \bibnamefont{Joachain}},
  \emph{\bibinfo{title}{Physics of Atoms and Molecules}}
  (\bibinfo{publisher}{Longman}, \bibinfo{address}{London and New York},
  \bibinfo{year}{1989}).

\bibitem[{\citenamefont{Marion and Thornton}(1988)}]{Marion}
\bibinfo{author}{\bibfnamefont{J.~B.} \bibnamefont{Marion}} \bibnamefont{and}
  \bibinfo{author}{\bibfnamefont{S.~T.} \bibnamefont{Thornton}},
  \emph{\bibinfo{title}{Classical dynamics}} (\bibinfo{publisher}{Harcourt
  Brace Jovanovich}, \bibinfo{address}{London, New York, Toronto},
  \bibinfo{year}{1988}).

\end{thebibliography}

\end{document}